\documentclass[11pt,draftcls,onecolumn]{IEEEtran}
\usepackage{cite}
\usepackage{amsmath,graphicx,amssymb}
\usepackage{MnSymbol}
\usepackage{wasysym}

\newtheorem{thm}{Theorem}

\newtheorem{rem}{Remarks}
\newtheorem{coy}{Corollary}

\makeatletter
\newcommand{\rom}[1]{\lowercase\expandafter{\romannumeral #1\relax}}
\makeatother

\begin{document}
%\title{Space-Time-Frequency Degrees of Freedom of $3D$ Multipath Wireless Channels: Intrinsic Limits for Spatial Information}
%\title{Degrees of freedom of bandlimited signals observed over finite time and space windows: An Upper Bound}
\title{Band Limited Signals Observed Over Finite Spatial and Temporal Windows: An Upper Bound to Signal Degrees of Freedom}


\author{\IEEEauthorblockN{Farhana Bashar, S.M. Akramus Salehin and Thushara D. Abhayapala}\\
\IEEEauthorblockA{Research School of Engineering\\
The Australian National University (ANU) \\
Canberra 0200 ACT, Australia\\
\{farhana.bashar,akramus.salehin,thushara.abhayapala\}@anu.edu.au}}

\maketitle

\begin{abstract}
The study of degrees of freedom of signals observed within spatially diverse
broadband multipath fields is an area of ongoing investigation and has a wide range of applications, including characterising
broadband MIMO and cooperative networks. However,
a fundamental question arises: given a size limitation on the observation
region, what is the upper bound on the degrees of
freedom of signals observed within a broadband multipath field over a
finite time window? In order to address this question, we characterize the multipath
field as a sum of a finite number of orthogonal waveforms or spatial modes.
We show that (\rom{1}) the ``effective observation time" is independent of spatial modes
and different from actual observation time, (\rom{2}) in wideband transmission regimes,
the ``effective bandwidth" is spatial mode dependent and varies from the given frequency bandwidth.
These findings clearly indicate the strong coupling between space and time as well as space and frequency
in spatially diverse wideband multipath fields. As a result, signal degrees of freedom
does not agree with the well-established degrees of freedom result as a product of spatial degrees of freedom and time-frequency degrees of freedom. Instead, analogous to Shannon's communication model where signals are
encoded in only one spatial mode, the available signal degrees
of freedom in spatially diverse wideband multipath fields is the {\em time-bandwidth}
product result extended from one spatial mode to finite
modes. We also show that the degrees of freedom is affected by the acceptable signal to noise ratio (SNR) in each spatial mode.
\end{abstract}


\section{Introduction}

\subsection{Motivation and Background}
In wireless communications, information is transmitted in the form
of waves and space is considered as the physical medium for
information transfer. Hence, as a physical process, waves propagate
in space via line of sight or multiple paths due to reflection,
diffraction and scattering by objects present in the physical
environment. Like any other physical phenomenon, wave propagation is
governed by the laws of physics. These laws determine the process
itself as well as the amount of diversity waves carry along their
path. The spatial diversity of multipath influences the amount of
information that can be communicated through wave propagation, thus,
using spatial diversity of multipath we can ensure better system
performances, including capacity improvement, high transmission
rate, improved bit error rate etc., \cite{PaulrajPapadias,Kohno}. In
effect, the study of the spatial degrees of freedom of different
multiple antenna systems (i.e., multi-user MIMO systems, distributed
MIMO systems, MIMO cognitive radio systems etc.) has gained renewed
attention and has more recently been addressed by
\cite{YairGoldsmith12,HuaGengGouJafar12,AyferOlivierTse12}. This
motivates to study the fundamental limits that space imposes on the
degrees of freedom of band limited signals observed over finite
spatial and temporal windows.

In this paper, our aim is to determine the upper limit to the
degrees of freedom of signals available in a band limited multipath
wavefield when the wavefield is observed in, or coupled to a limited
source-free region of space over a finite time window. We may
assume that multiple antennas or sensors are located in the region
of space to sample the observable multipath field for signal
processing or communication purposes. We, however, aim to find an upper bound on the available degrees of freedom without
explicitly considering a specific propagation condition, physical
setup or application and thus, to show that the coupling of time and
band limited multipath signals into a spatial region is
fundamentally limited by a finite number of spatial modes.
Throughout the paper, we will frequently refer to the radius/ size of
the $3D$ multipath observation region\footnote{In antenna propagation and
sensor array signal processing applications, an alternative
terminology is the effective antenna aperture.}. Our derived result has great significance in a wide range
of applications, including (\rom{1}) measuring the number of receive antennas required to sample
a given region to maximize the performance gain, (\rom{2}) characterizing broadband beamforming techniques for next generation
wireless communication to provide high quality video and audio, (\rom{3}) developing interference alignment scheme for MIMO wireless networks,
(\rom{4}) characterizing the degrees of freedom of distributed multi-antenna communications for broadband transmissions.

We review the degrees of freedom available in spatially diverse
multipath fields in different contexts. Earlier works,
\cite{telatar,Foschini} focused on multipath fields that exhibit
rich scattering and there are independent fading paths between
transmitter and receiver antenna elements. According to these works,
available degrees of freedom is the minimum number of transmit and
receive antenna elements and channel capacity can be improvement
remarkably by increasing number of the antenna elements. However,
insufficient antenna spacing violates the assumption of independent
fading and prevent channel capacity to increase linearly with
degrees of freedom \cite{TealAbhayapalaKennedy}. The impact of fading correlation on
spatially diverse multipath fields was studied by a large number of
research works (e.g., \cite{sayeed02,Liuetal03}). Afterwards,
independent works \cite{Miller00,pollokTDrod031,hanlenFu} provided
the characterization of the spatial degrees of freedom in
multi-antenna systems as a function of the area, geometry of the
antenna arrays and the angular spread of the physical environment.
In addition, \cite{rodTDjones,rodParastooTDjones} estimated the
degrees of freedom available in source-free narrowband multipath
fields observed over a spatial window and showed that the
available degrees of freedom scales with the spatial dimension in
terms of wavelengths. In contrast, Poon et al.
\cite{PoonBrodersenTse} and Franceschetti \cite{Franceschetti2012}
applied antenna theory and Slepian's theory of spectral
concentration, respectively, to derive a fundamental limit on the
degrees of freedom available in a wideband multi-antenna systems for
a given constraint on the area of the spatial region and observation
time and defined the degrees of freedom as a product of spatial
degrees of freedom and degrees of freedom of the wideband channel
itself. Since for wideband transmissions, space time, and frequency
are strongly coupled, available bandwidth and observation time over
space respectively differ from actual bandwidth and observation time
depending on the available spatial information, the works of
\cite{PoonBrodersenTse, Franceschetti2012} did not take this into
account. In another approach, \cite{FranceschettiChakraborty09}
characterized multi-antenna systems in a  wideband transmission
regime and stated that in case of wideband frequency transmission,
space and time are strongly coupled. However, how information is
conserved in space-time was left as an open and important problem.

\subsection{Our Approach and Contributions}

The analysis in this paper considers a wideband multipath wavefield
observed within a limited source-free region of space over a finite
time window. The signals observable within this wavefield are
studied as solutions to the Helmholtz wave equation
\cite{coltonkress98} and they are encoded in infinite but countable
number of orthogonal waveforms or spatial modes. This mathematical
framework is similarly used in \cite{TDLeif1,hanlenTD}. However, in
comparison, our derived result is more accurate, since we have
considered the affect of available spatial information not only on
the frequency bandwidth but also on the observation time. Further,
the degrees of freedom result provided in \cite{TDLeif1,hanlenTD} is
derived by using a complex geometrical argument to extend the
narrowband degrees of freedom result of \cite{rodTDjones} to a
broadband scenario and resulted in a loose bound. Further, it is
unclear, for different spatial modes, how the usable (effective)
bandwidth varies from the given frequency bandwidth. In this work,
on the contrary, the degrees of freedom result is derived in a
simple manner. Moreover, we clarify that at each spatial mode, how
(and why) the observable signals are band limited within an
effective frequency bandwidth rather than the given frequency
bandwidth. In addition, we illustrate that beyond a certain spatial
mode, the effective bandwidth becomes zero which in turn, truncates
the wavefield from its infinite representation to a
finite number of spatial modes. Afterwards, by counting the number
of spatial modes required to represent any signal within the given
multipath field, we derive an analytical expression to determine the
degrees of freedom of the signal\footnote{A preliminary study for the degrees of freedom were presented previously in \cite{BasharSalehinAbhayapala14} and \cite{fbashar_12(1)}, respectively, for $2D$ wavefields and $3D$ wavefields.}.

We depict the strong coupling relation between space
and time as well as space and frequency in spatially diverse
wideband multipath fields. We show that the effective observation
time is fixed, independent of spatial modes, different from given
observation time and essentially related to the spatial dimension of
the observable region. Whereas, for broadband transmissions, at each spatial mode,
the observable signal is band limited within an effective frequency bandwidth, since
even though the usable bandwidth at the lower spatial modes
is equal to the given frequency bandwidth, for the higher modes, the usable bandwidth
is less than the given frequency bandwidth. The
coupling relations also indicate that for spatially diverse wideband
multipath fields, the classical degrees of freedom result of {\em
time-bandwidth product} can not be extended directly to the product of
spatial degrees of freedom and {\em
time-bandwidth product} as shown in \cite{PoonBrodersenTse,Franceschetti2012}, rather the degrees of
freedom result should portray how the time and band limited signals
are coupled to a limited region of space. Our derived degrees of
freedom result evidently portrays the impact of the coupling
relations on the available degrees of freedom in spatially diverse
wideband multipath fields. We also show the affect of the acceptable
signal to noise ratio (SNR) on the available degrees of
freedom of each spatial mode .

\subsection{Organization}
The reminder of the paper is organized as follows. In Section
\ref{problem}, the problem statement together with background on Shannon's time-frequency degrees of freedom and the eigenbasis expansion of the wavefield are discussed. In Section \ref{results}, we
present our main results, while, Section \ref{plot} provides graphical
analysis of our derived results. Next, Section \ref{insights}
elaborates the physical insights of the main results and briefly
discusses the applications. We summarize the main contributions of this paper in Section \ref{conclusion}.

\section{Problem Statement and Background}\label{problem}

\subsection{Physical Problem}

In this paper, we consider a multipath field band limited to
$[F_0-W, F_0+W]$ and observed over a time window $[0, T]$ within a
$3D$ spatial window enclosed by a spherical region of radius $R$.
Here, $F_0$ represents the mid band frequency. Any signal sampled
or recorded within this spatial region can be
expressed as a function of space and time whose spectra lies within
the frequency range and whose time function lies within the time
interval. Since it is not possible to confine any waveform in both
time and frequency, we consider that the spectrum is confined
entirely within the frequency range and the time function is
negligible outside the
time interval. \\
We now express the space-time signal as
\begin{IEEEeqnarray}{rCl}
\psi(\textbf{\emph{x}},t) =
\frac{1}{2\pi}\int^{\infty}_{-\infty}\Psi(\textbf{\emph{x}},\omega)e^{j
\omega t} d\omega
\end{IEEEeqnarray}
%%%%%
where $\Psi(\textbf{\emph{x}},\omega)$ is the Fourier transform of
$\psi(\textbf{\emph{x}},t)$ with respect to $t$, \textbf{\emph{x}}
represents a position in 3D space, such that
$r=\|\textbf{\emph{x}}\|\leq R$ denotes the euclidean distance of
\textbf{\emph{x}} from the origin, which is the center of the region
of interest and $j=\sqrt-1$. Due to the band limitedness,
$\Psi(\textbf{\emph{x}},\omega)$ is assumed to be zero outside the
band $[F_0-W,F_0+W]$. Thus, the space-time signal can be rewritten
as
\begin{IEEEeqnarray}{rCl} \label{timespace}
\psi(\textbf{\emph{x}},t) = \frac{1}{2\pi}\int^{2\pi (F_0+W)}_{2\pi
(F_0-W)}\Psi(\textbf{\emph{x}},\omega)e^{j\omega t} d\omega.
\end{IEEEeqnarray}

In this work, we aim to answer the fundamental question: {\em Given
constraints on the size of the observation region, frequency bandwidth
and observation time window, what is the upper bound on the degrees of
freedom of signals observable within multipath wireless fields?}

We will utilize Shannon's result \cite{shannon49} that provides the degrees of
freedom of temporal signals for band limited communications over a
single channel. Shannon's result states that if the transmission is
band limited to $[-W, W]$ and limited to the time interval
$[0,T]$, the available time-frequency degrees of freedom is limited to
$2WT+1$. We provide more detailed reasoning of Shannon's result
in the next subsection.

\subsection{Shannon's Time-Bandwidth Product} \label{shannon}
Let us now review the reasoning behind the time-bandwidth product
result \cite{shannon49}. In time domain we have a wideband signal
and in frequency domain this signal can be expressed as a spectrum.
The mapping between these two domains is the Fourier transform. The
spectrum is then expanded over the frequency range with the help of
Fourier series expansion. This expansion represents the time domain
signal by a weighted sum of orthogonal basis functions. Given the
signal is approximately time limited to $[0,T]$ and its spectrum is
band limited to $[-W, W]$, the minimum number of terms required in
the sum to satisfy both of these constraints provide the available
degrees of freedom of the signal, $2WT+1$. Note that there may be
slight discrepancy as the time domain signal obtained by the Fourier
series expansion over the time interval will not be strictly limited
within the frequency band, rather it may contain some frequency
component outside the band. However, in another approach,
\cite{SlepianPollak,LandauPollak1,LandauPollak2,Slepian83} argued
that roughly $2WT+1$ samples are enough to approximate any signal in
energy for the best choice of a complete set of band limited
functions which possess the property of being orthogonal over a
given finite time interval. Afterwards, the time-bandwidth product
result was formalized by several authors
\cite{dollard,bennett,jerri} for various other configurations.

However, Shannon's result only accounts for broadband transmission
over a single channel. In multipath wireless fields sampled over a
region of space, spatial diversity is exploited, for instance, in
MIMO communications, providing several independent channels over
which information can be transmitted. To account for the spatial
diversity of wireless fields, we start with the spherical harmonics
analysis of the wavefield observed within a region of space.

\subsection{Spherical Harmonics Analysis of Wavefields}

We consider the space-frequency spectrum $\Psi(\textbf{\emph{x}},\omega)$ in \eqref{timespace} as
a scalar wavefield observed within a $3D$ spherical region of finite
radius $R$ generated by a source or distribution of sources and
scatterers that exist outside the region of interest at some radius
$R_s> R$. Hence, $\Psi(\textbf{\emph{x}},\omega)$ satisfies the
Helmholtz wave equation (in the region of interest)
\cite{ArfkenWeber}
\begin{IEEEeqnarray}{rCl} \label{helmholtz}
\nabla^2 \Psi(\textbf{\emph{x}},\omega) + k^2
\Psi(\textbf{\emph{x}},\omega) =0
\end{IEEEeqnarray}
where $\nabla^2$ is the Laplacian, $k= \omega / c$ is the scalar
wavenumber, $c$ is the wave velocity and $\omega$ is the angular
frequency which can be expressed in terms of usual frequency $f$ as
$\omega= 2\pi f$. Note that even though we only consider scalar
waves, our derived results are equally valid for vector waves
\cite[pp. 166]{AndersGerhardStaffan}.

In the spherical coordinate system, the wavefield
$\Psi(\textbf{\emph{x}},\omega)$ in \eqref{helmholtz} can be
decomposed into spherical harmonics which form an orthogonal basis
set for the representation of the wavefield. Using the Jacobi-Anger
expansion \cite[pp. 32]{coltonkress98} and the addition theorem
\cite[Theorem 2.8]{coltonkress98}, we can express
$\Psi(\textbf{\emph{x}},\omega)$ as follows
%%%%%%%%
\begin{IEEEeqnarray}{rCl} \label{expansion}
\Psi(\textbf{\textit{x}}, \omega) = \sum_{n=0}^{\infty}
\sum_{m=-n}^{n} \Psi_{nm}(r, \omega) Y_{nm}(\hat{\textbf{\emph{x}}})
\end{IEEEeqnarray}
%%%%%%%%
where spatial mode $n (\geq 0)$ and spatial order $m$  ($ |m| \leq n$) are integers,
such that for any particular mode $n$, there are $2n+1$ orders,
$\hat{\textbf{\emph{x}}}\triangleq
\textbf{\emph{x}}/\|\textbf{\emph{x}}\|$ is the unit vector in the
direction of nonzero vector \textbf{\emph{x}}, $Y_{nm}(\cdot)$ are
the spherical harmonics and $\Psi_{nm}(r, \omega)$ are the harmonic
coefficients. These harmonic coefficients can be expressed as the
product of the frequency dependent coefficients
$\alpha_{nm}(\omega)$ and the spherical Bessel functions of the
first kind $j_n(\omega r/c)$ \cite[p. 227]{EGWilliams} as
%%%%%%%%%%%%%
\begin{IEEEeqnarray}{rCl} \label{signal}
\Psi_{nm}(r, \omega)\triangleq \alpha_{nm}(\omega)
j_n\left(\frac{\omega}{c}r\right).
\end{IEEEeqnarray}
%%%%%%%%%
Further, we can think of $\Psi_{nm}(r, \omega)$ as the space-frequency
spectrum encoded in $n$ modes and $m$ orders. In this work, we frequently refer to
$(m,n)^{th}$ mode space-frequency spectrum which represents the spectrum at a particular mode $n$ and order $m$.
On the contrary, $n^{th}$ mode space-frequency spectrum refers to the $n^{th}$ mode spectrum considering all
of the $2n+1$ orders.\\
Note that $\Psi_{nm}(r, \omega)$ depends only on frequency and
radial coordinate $r$ of vector $\textbf{\emph{x}}$, not on angular
information of vector $\textbf{\emph{x}}$. Also note that the
spherical harmonics $Y_{nm}(\cdot)$ exhibit the following
orthonormal property \cite[p. 191]{EGWilliams}
\begin{IEEEeqnarray}{rCl} \label{orthogonal}
\int_{\Omega}
Y_{nm}(\hat{\textbf{\emph{x}}})Y^{\ast}_{\acute{n}\acute{m}}(\hat{\textbf{\emph{x}}})
d\Omega= \delta_{n\acute{n}} \delta_{m\acute{m}}
\end{IEEEeqnarray}
where the integration is taken over the unit sphere $\Omega$ and
$\delta_{nm}$ is the Kronecker delta function which is defined as
$\delta_{nn}=1$ and $\delta_{nm}=0$ if $n\neq m$.\\
The expansion of the wavefield \eqref{expansion} can be viewed as a
weighted sum of orthogonal spherical waveforms encoded in an
infinite but countable number of spatial modes. This expansion was
used in \cite{rodTDjones,rodParastooTDjones} to represent general
narrowband multipath fields observed over a region of space. The
work of \cite{rodTDjones,rodParastooTDjones} truncated the spherical
harmonic expansion of the wavefield to a finite number of spatial
modes which contain most of the energy. The number of spatial modes
in the truncated expansion indicate the spatial degrees of freedom
(i.e., the number of independent channels). Hence, in this work, we
need to find a suitable way to truncate this expansion for broadband
multipath fields observed over a region of space.

\section{Main Results} \label{results}

In this section, we derive an upper bound to the available degrees
of freedom of any signal observed within a band limited multipath field
over a spherical region of finite radius for a finite time interval.
Before proceeding to the main result presented in this work, we
can reasonably ask, how the observable time and band limited signals are
coupled to a limited region of space for each spatial mode?
In order to answer this question, we first show the coupling of time
limited signals to a finite spatial window for each spatial mode $n$.

\subsection{Effective Observation Time of the Spatial Modes}
Lets consider the multipath wavefield is generated by a single farfield source transmitting a time domain signal. The wavefield is enclosed within a spherical region of radius $R$. Hence, time required for the time domain signal to travel across the diameter of the spherical region is $2R/c$. Further, the wavefield is observed over a time window $[0, T]$. As a result, the observable multipath field captures information content of the time domain signal over a time interval $T+2R/c$. In the following theorem, we formalize this statement for the $n^{th}$ mode space-time wavefield $\psi_{nm}(r,t)$ generated by the $n^{th}$ mode time domain signal $a_{nm}(t)$.
\begin{thm} [Observation time of the spatial Modes]
Given a multipath field observed over a spherical region of radius $R$
for a time interval $T$ that is encoded in a countable number of
spatial modes $n$, then it is possible to capture information about the underlying $n^{th}$
mode time domain signal $a_{nm}(t)$ over an effective time interval
\begin{eqnarray}\label{TS}
T_{eff}=T+\frac{2R}{c}.
\end{eqnarray}%}
Further, this effective time interval $T_{eff}$ is independent of
the spatial mode index $n$.
\end{thm}
Proof of the theorem is provided in Appendix \ref{A1}.

\begin{rem}
Observing the $n^{th}$ mode space-time wavefield $\psi_{nm}(r, t)$ over a time window $[0, T]$ within a spatial window $0 \leq r \leq R$ is equivalent to observing the $n^{th}$ mode time domain signal $a_{nm}(t)$ over the time window $[-R/c, T+R/c]$. Hence, the effective time interval is essentially related to the spatial observation region. This indicates the coupling relation between space and time. Further, the effective time interval is fixed and independent of spatial modes $n$.
\end{rem}

In the next subsection, we show the coupling relation between space and frequency.
This relation truncates the expansion in \eqref{expansion}
to a finite number of spatial modes.

\subsection{Effective Bandwidth of the Spatial Modes} \label{effectiveBW}

The performance of wireless communication systems is highly
determined by noise. Ideally, if the wireless communication systems
are noiseless, it would be possible to measure signals with infinite
precision and each spatial mode $n$ would have an effective
bandwidth equal to the given frequency bandwidth, i.e., from $F_0-W$
to $F_0+W$. However, in practical systems, signals are perturbed by
noise. Hence, it is not possible to detect signals within the band
of frequencies where the signal to noise ratio (SNR) drops below a
certain threshold $\gamma$. This threshold is dependent on the
antenna/ sensor sensitivity or the robustness of the signal
processing method to noise.

To determine how noise affects the available bandwidth at each
spatial mode $n$, let us assume that
$\eta_R(\hat{\textbf{\emph{x}}}, \omega)$ is the white Gaussian
noise on the surface of the spherical region (at radius $R$)
associated with the antenna/ sensor at the angular position
$\hat{\textbf{\emph{x}}}$. Hence, from \eqref{expansion} and \eqref{signal}, the space-frequency spectrum on
the sphere is
\begin{IEEEeqnarray} {rCl} \label{field}
\Psi(R, \hat{\textbf{\emph{x}}},\omega) \negthickspace = \negthickspace \sum_{n=0}^{\infty}
\sum_{m=-n}^{n} \negthickspace \alpha_{nm}(\omega)
j_n\left(\frac{\omega}{c}r\right)\negthickspace Y_{nm}(\hat{\textbf{\emph{x}}})\negthickspace + \negthickspace
\eta_R(\hat{\textbf{\emph{x}}}, \omega).
\end{IEEEeqnarray}
In the following theorem and corollary, we characterize the white Gaussian
noise at the different modes.

\begin{thm} [White Gaussian Noise in $L^2$] Given a zero
mean white Gaussian noise with variance $\sigma_0 ^2$ in $L^2
(\mathbb{S}^2)$ represented by a random variable
$\eta_R(\hat{\textbf{\emph{x}}})$ where $\hat{\textbf{\emph{x}}} \in
\mathbb{S}^2$, such that for any function
$\psi_i(\hat{\textbf{\emph{x}}})\in L^2(\mathbb{S}^2)$ the complex
scalar
\begin{IEEEeqnarray} {rCl}
\nu_i \triangleq \int_{\mathbb{S}^2} \eta_R(\hat{\textbf{\emph{x}}})
\psi_i^\ast (\hat{\textbf{\emph{x}}}) d\hat{\textbf{\emph{x}}} =
\langle \eta_R(\hat{\textbf{\emph{x}}}),
\psi_i(\hat{\textbf{\emph{x}}})\rangle
\end{IEEEeqnarray}
is also a zero mean Gaussian random variable with variance
$E\{{|\nu_i|}^2 \}= \sigma_0^2  \int_{\mathbb{S}^2}
{|\psi_i(\hat{\textbf{\emph{x}}})|}^2 d\hat{\textbf{\emph{x}}}=
\sigma_0^2 ({\|\psi_i(\hat{\textbf{\emph{x}}})\|}_{L^2})^2$.
\cite[eqn 8.1.35]{gallager}
\end{thm}

\begin{figure} [ht]
\centering
\includegraphics[width=\columnwidth]{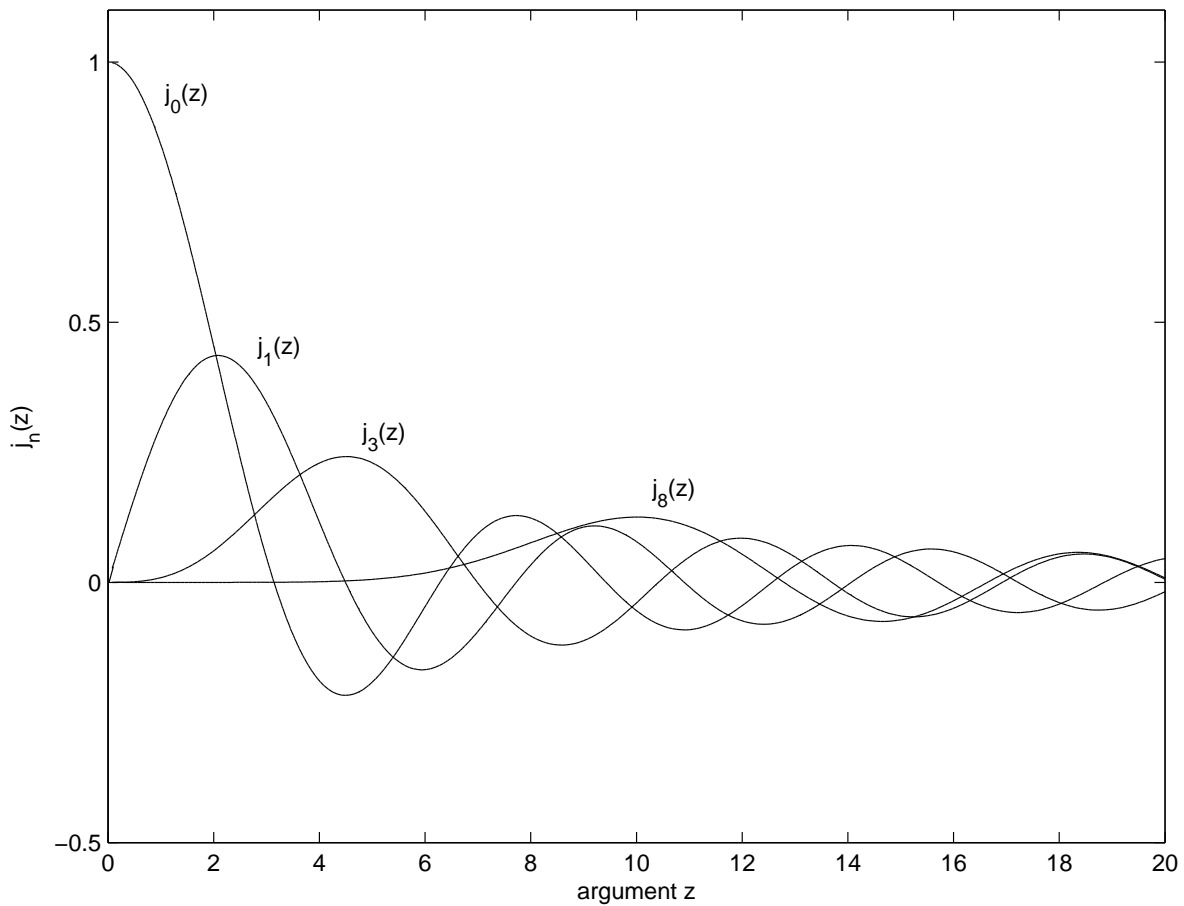}\\
\caption{ Spherical Bessel functions of first kind $j_{n}(z)$ vs.
argument $z$ for $n= 0, 1, 3, 8$.} \label{sbessel}
\end{figure}

\begin{coy} Considering $\psi_i(\hat{x})$
to be the orthonormal basis functions
$Y_{nm}(\hat{\textbf{\emph{x}}})$, the spatial Fourier coefficients
for the noise is
\begin{IEEEeqnarray} {rCl} \label{noise}
\nu_{nm}(\omega) = \int_{\mathbb{S}^2}
\eta_R(\hat{\textbf{\emph{x}}}, \omega) Y_{nm}^\ast
(\hat{\textbf{\emph{x}}}) d\hat{\textbf{\emph{x}}}.
\end{IEEEeqnarray}
Applying Theorem $2$, $\nu_{nm}(\omega)$ are also zero mean Gaussian
random variables with variance
\begin{IEEEeqnarray} {rCl}
E\{{|\nu_{nm}(\omega)|}^2 \}= \sigma_0^2 (\omega) \int_{\mathbb{S}^2}
{|Y_{nm}^\ast
(\hat{\textbf{\emph{x}}})|}^2 d\hat{\textbf{\emph{x}}}=
\sigma_0^2(\omega)
\end{IEEEeqnarray}
where the noise power is independent of the mode. Further, since the noise is white Gaussian,
the noise power is the same at all frequencies $\omega$, i.e.,
\begin{IEEEeqnarray} {rCl}
E\{{|\nu_{nm}(\omega)|}^2 \}= \sigma_0^2.
\end{IEEEeqnarray}

\end{coy}
Based on Corollary $1$, we can define the $(m,n)^{th}$ mode space-frequency
spectrum at radius $R$ as
\begin{IEEEeqnarray} {rCl} \label{coefficient}
\varPsi_{nm}(R,\omega) = \alpha_{nm}(\omega)
j_n\left(\frac{\omega}{c}R\right)+ \nu_{nm}(\omega)
\end{IEEEeqnarray}
and we assume that the noise and the signal are not dependent on
each other. Here, the $(m,n)^{th}$ mode space-frequency spectrum
$\varPsi_{nm}(R,\omega)$ takes white Gaussian noise
$\nu_{nm}(\omega)$ into account. This white Gaussian noise has the property that each spatial mode is
perturbed independently of all the others. Further, $\alpha_{nm}(\omega)$ is
the $(m,n)^{th}$ mode signal spectrum band limited over the range
$[F_0-W, F_0+W]$. In contrast, for a fixed value of the radius,
$j_n(\omega R/c)$ can be treated as a function of frequency.
However, it is evident from Fig. \ref{sbessel} that except for the
the $0^{th}$ order, the spherical Bessel functions $j_n(z)$ start
small before increasing monotonically to their maximum. Therefore,
for frequencies less than a critical frequency $F_n$, the magnitude
of the $n^{th}$ order spherical Bessel function $|j_n(\omega R/c)|$
is negligible.
\\
At each spatial mode $n>0$, for frequencies less than a
critical frequency $F_n$, the SNR is less than the threshold
$\gamma$. As a result, we can not detect the space-frequency spectrum for
frequencies less than $F_n$. In addition, as depicted in Fig.
\ref{sbessel}, the spherical Bessel functions $j_n(z) (n>0)$ start
more slowly as $n$ increases. Thus, after a certain spatial mode $n$,
the critical frequency $F_n$ is larger than the lower bound of the
given bandwidth $F_0-W$ and the useable (effective) bandwidth of
this mode and the modes above varies from the given frequency
bandwidth. The following theorem provides the effective bandwidth available at
each spatial mode $n$.
\begin{thm}[Effective Bandwidth of the $n^{th}$ Mode]
Any wireless multipath wavefield observed within a spherical region
of space bounded by radius $R$ over a frequency range
$[F_0-W,F_0+W]$ is encoded in a finite number of spatial modes $0
\leq n\leq N_{\max}$ where the effective frequency bandwidth of the
$n^{th}$ mode space-frequency spectrum $\varPsi_{nm}(R,\omega)$ is
\begin{IEEEeqnarray} {rCl}\label{BW}
W_n = \begin{cases} 2W, & \mbox{$0\leq n\leq N_{\min}$} \\
F_0\negthickspace+\negthickspace
W- F_n, & \mbox{$N_{\min}<n\leq N_{\max}$} \\
0, & \mbox{\text{otherwise.}} \end{cases}
\end{IEEEeqnarray}\\
Here, $N_{\min}$ is the lowest spatial mode beyond which the
critical frequency $F_n > F_0-W$, $N_{\max}$ is the lowest spatial
mode beyond which the critical frequency $F_n> F_0+W$ and
\begin{IEEEeqnarray} {rCl} \label{cf}
F_n = \frac{nc}{e \pi R} + \frac{c}{2 e \pi R} \log \left(
\frac{\gamma}{{(SNR)}_{\alpha,\max}} \right)
\end{IEEEeqnarray}
with the threshold $\gamma$ depicting the ability of the system to
detect signals buried in noise. Further, we consider that the power
of the $n^{th}$ mode signal spectrum $\alpha_{nm}(\omega)$ is finite
and bounded for all modes $n$, orders $m$ and frequencies $\omega$,
i.e.,
\begin{IEEEeqnarray} {rCl} \label{emax}
E\{{|\alpha_{nm}(\omega)|}^2\}\leq {|\alpha_{\max}|}^2,~~
\mbox{$\forall n,m,\omega$}
\end{IEEEeqnarray}
hence, the maximum SNR of the signal spectrum $\alpha_{nm}(\omega)$
for any mode $n$ is
\begin{IEEEeqnarray} {rCl} \label{k}
{(SNR)}_{\alpha,\max} = \frac{{|\alpha_{\max}|}^2}{\sigma_0 ^2}.
\end{IEEEeqnarray}
\end{thm}
Proof of the theorem is given in Appendix \ref{A2}.

\begin{rem} The effective frequency bandwidth of
each spatial mode is related to the spatial observation region and
varies from the given frequency bandwidth depending on the critical
frequency $F_n$. This portrays the strong coupling relation between
space and frequency. Further, for $n>N_{\max}$, the critical
frequency $F_n$ is greater than the upper bound of the given
frequency range $F_0+W$. Therefore, we can truncate the expansion in
\eqref{expansion} to a finite number of spatial modes as
\begin{IEEEeqnarray}{rCl} \label{expansion1}
\Psi(\textbf{\textit{x}}, \omega) = \sum_{n=0}^{N_{\max}}
\sum_{m=-n}^{n} \varPsi_{nm}(r, \omega)
Y_{nm}(\hat{\textbf{\emph{x}}}).
\end{IEEEeqnarray}
Using \eqref{cf}, the upper bound for the spatial modes $N_{\max}$ is
\begin{IEEEeqnarray}{rCl} \label{nmax}
N_{\max}= \left\lceil e \pi (F_0+W) \frac{R}{c} + \frac{1}{2}
\log\left(\frac{{(SNR)}_{\alpha,\max}}{\gamma}\right) \right\rceil
\end{IEEEeqnarray}
where $\lceil \cdot \rceil$ is the ceiling value, since by
definition spatial modes are integers.
\end{rem}
\subsection{Upper Bound to Signal Degrees of Freedom}

We are now in a position to provide an upper bound to the available
degrees of freedom of wideband signals observed over finite spatial
and temporal windows. In order to do so, it is useful to think of
$(m,n)^{th}$ mode space-frequency spectrum $\varPsi_{nm}(r, \omega)$
in time domain, in which case we obtain
\begin{IEEEeqnarray}{rCl}\label{s}
\psi_{nm}(r,t)= \frac{1}{2\pi}\int_{\Omega_n} \varPsi_{nm}(r,
\omega) e^{j\omega t} d\omega
\end{IEEEeqnarray}
where $\psi_{nm}(r,t)$ is the inverse Fourier transform of
$\varPsi_{nm}(r, \omega)$ with respect to $\omega$ and the
integration is taken over $\Omega_n$ with $\Omega_n\in [2\pi
(F_0-W), 2\pi (F_0+W)]$ for $0\leq n\leq N_{\min}$ and $\Omega_n\in
[2\pi F_n,
2\pi (F_0+W)]$ for $N_{\min}<n\leq N_{\max}$ where $F_n$ is defined in \eqref{cf}. \\
We expand $\varPsi_{nm}(r,\omega)$ over the frequency range using
the Fourier series expansion, similar to \cite{shannon49}, as follows
\begin{IEEEeqnarray}{rCl} \label{S}
\varPsi_{nm}(r,\omega)= \sum_{\ell=-\infty}^{\infty} c_{nm\ell}(r)
e^{-j\omega \frac {\ell}{W_n}}
\end{IEEEeqnarray}
where the Fourier coefficients
\begin{align}\label{c}
c_{nm\ell}(r)&= \frac{1}{2\pi W_n} \int_{\Omega_n}
\varPsi_{nm}(r, \omega) e^{j\omega \frac {\ell}{W_n}}d\omega  \nonumber \\
&= \frac {1}{W_n} \psi_{nm}(r,\frac{\ell}{W_n})
\end{align}
are proportional to the samples of $\psi_{nm}(r,t)$ and $W_n$ is the
effective frequency of the $n^{th}$ mode defined in \eqref{BW}. The
result \eqref{c} is obtained from \eqref{s} when $t=\ell/W_n$. It
illustrates that the samples of $\psi_{nm}(r,t)$ determine the
coefficients $c_{nm\ell}(r)$ in the Fourier series expansion.
Therefore, analogous to Shannon's work \cite{shannon49}, we can
reconstruct the $(m,n)^{th}$ mode space-time signal $\psi_{nm}(r,t)$
from its samples as follows
\begin{IEEEeqnarray}{rCl} \label{ftruncation}
\psi_{nm}(r,t)&=&
\sum_{\ell=-\infty}^{\infty}\psi_{nm}(r,\frac{\ell}{W_n})e^{j2\pi
W_{0n}(t-\frac{\ell}{W_n})} \frac{\sin\pi
W_n(t-\frac{\ell}{W_n})}{\pi W_n(t-\frac{\ell}{W_n})}
\end{IEEEeqnarray}
where $W_{0n}$ is the mid band frequency of the $(m,n)^{th}$ mode.
We obtain \eqref{ftruncation} by substituting the Fourier series
\eqref{S} in \eqref{s}, applying \eqref{c} and then exchanging
integration and
summation.\\
Hence, it is possible to
reconstruct the space-time signal $\psi(\textbf{\emph{x}},t)$
\eqref{timespace} by summing the $(m,n)^{th}$ mode space-time
signals for all possible values of $n$ and $m$ over all propagation
directions, i.e.,
\begin{IEEEeqnarray}{rCl}
\psi(\textbf{\emph{x}},t) &=& \sum_{n=0}^{N_{\max}} \sum_{m=-n}^n
\psi_{nm}(r,t) Y_{nm}(\hat{\textbf{\emph{x}}})
\end{IEEEeqnarray}
and substituting \eqref{ftruncation} yields
\begin{IEEEeqnarray}{rCl} \label{STsignal}
\psi(\textbf{\emph{x}},t) &=& \sum_{n=0}^{N_{\max}} \sum_{m=-n}^n
\sum_{\ell=-\infty}^{\infty}\psi_{nm}\left(r,\frac{\ell}{W_n}\right)e^{j2\pi
W_{0n}(t-\frac{\ell}{W_n})} \frac{\sin\pi
W_n(t-\frac{\ell}{W_n})}{\pi
W_n(t-\frac{\ell}{W_n})}Y_{nm}(\hat{\textbf{\emph{x}}}).
\end{IEEEeqnarray}
Observe that the spherical harmonics $Y_{nm}(\hat{\textbf{\emph{x}}})$ are orthogonal over the spherical region as shown in \eqref{orthogonal}.
Further, considering
\begin{IEEEeqnarray}{rCl}
\phi_{\ell}(t) =e^{j2\pi
W_{0n}(t-\frac{\ell}{W_n})} \frac{\sin\pi
W_n(t-\frac{\ell}{W_n})}{\pi
W_n(t-\frac{\ell}{W_n})},
\end{IEEEeqnarray}
the functions $\phi_{\ell}(t)$ are orthogonal over time. Proof of the orthogonality of
the functions $\phi_{\ell}(t)$ is provided in Appendix \ref{A3}. Therefore, following the same reasoning as Shannon \cite{shannon49}, discussed in
Section \ref{shannon}, the minimum numbers of terms required in the
sum \eqref{STsignal} that satisfy the constraints on observation region size, bandwidth
and observation time window provide the available signal degrees of freedom
within the given multipath field. Here, $\ell$ can be truncated to
$[0, W_nT_{eff}]$. We truncate $\ell$ based on the fact that the
$(m,n)^{th}$ mode space-time signal $\psi_{nm}(r,t)$
\eqref{ftruncation} is band limited to $W_n$. Hence, we can
determine $\psi_{nm}(r,t)$ by taking samples $1/W_n$ units apart.
Now, in order to limit $\psi_{nm}(r,t)$ within the interval
$T_{eff}$, $\psi_{nm}(r,\ell/W_n)$ is non-zero for only the
appropriate values of $\ell$, such that $0\leq \ell \leq
W_nT_{eff}$. This means that the degrees of freedom of the
$(m,n)^{th}$ mode space-time signal $\psi_{nm}(r,t)$ is
$1+W_nT_{eff}$. Hence, the total degrees of freedom of any signal
available in the given multipath field considering all modes and
orders is given by
%%%%%%
\begin{IEEEeqnarray}{rCl}
\label{eq:D} D= \sum_{n=0}^{N_{max}}\sum_{m=-n}^{n} (W_nT_{eff}+1)
\end{IEEEeqnarray}
%%%%%%
where for different values of $n$, $W_n$ is provided in \eqref{BW}
and $T_{eff}$ is defined in \eqref{TS}.

Using our previous results, we now provide the following theorem for
the degrees of freedom of any signal observed within a broadband multipath field.

\begin{thm} Given a multipath wireless field
band limited to $[F_0-W, F_0+W]$ over the time interval $[0, T]$
within a $3D$ spherical region of radius $R$, then the degrees of
freedom of any signal observable within this multipath field is
upper bounded by
\begin{IEEEeqnarray}{rCl} \label{Dspace}
D\leq {(N_{\max}+1)}^2 + 2W \left(T+\frac{2R}{c}\right) \Big[&& {(N_{\min}+1)}^2
\negthickspace + \negthickspace 2{\left(e\pi \frac{R}{c}\right)}^2
\left(F_0W-\frac{1}{3}W^2\right) \negthickspace +\negthickspace
e\pi\frac{R}{c}\negthickspace \left(2F_0
\negthickspace - \negthickspace W \right)\negthickspace \nonumber \\
&& + \log\left(\frac{{(SNR)}_{\alpha,\max}}{\gamma}\right)
\left(  e\pi W \frac{R}{c} +  1 \right)\Big]
\end{IEEEeqnarray}
where $N_{\max}$ is defined in \eqref{nmax} and from Theorem $3$,
$N_{\min}$ is the lowest spatial mode beyond which the critical
frequency $F_n > F_0-W$. Using \eqref{cf},
\begin{IEEEeqnarray}{rCl} \label{nmin}
N_{\min}= \left\lceil e \pi (F_0-W) \frac{R}{c} + \frac{1}{2}
\log\left(\frac{{(SNR)}_{\alpha,\max}}{\gamma}\right) \right\rceil.
\end{IEEEeqnarray}
Since by definition spatial modes are integers, $N_{\min}$ is
defined as a ceiling value.
\end{thm}
Proof of the theorem is given in Appendix \ref{A4}.

In the next section, we graphically illustrate our derived results.

\section{Graphical Illustrations} \label{plot}

Consider $\lambda_0$ as the wavelength corresponding to the mid band
frequency $F_0$. Therefore, $R$, $W$, and $T$  can be represented in
terms of $\lambda_0$ or $F_0$ as follows
\begin{eqnarray} \label{abd}
&R&= a \lambda_0, ~~~\mbox{$a\in[0,\infty)$} \nonumber \\
&W& = b F_0, ~~~\mbox{$b\in[0,1]$} \nonumber \\
&T&= d /F_0, ~~\mbox{$d\in[0,\infty)$}
\end{eqnarray}
where $a$, $b$ and $d$ are real numbers. Furthermore, $c = F_0
\lambda_0$ and $W = F_0$ represents the extreme broadband scenario. Hence, \eqref{Dspace} can be rewritten as
\begin{eqnarray} \label{newD}
D \leq (\eta_{\max}+ 1)^2 &+& b(2a+d)[2{(\eta_{\min}+ 1)}^2+ (2e\pi ab)^2(1/b-1/3) +2e\pi ab(2/b-1) \nonumber \\
&& \qquad \quad ~~~ + 2 \log \rho (e \pi ab +1)]
\end{eqnarray}
where considering $\rho =
{(SNR)}_{\alpha,\max}/\gamma$, $$\eta_{\max}= N_{\max}(a,b,\rho)=\lceil e\pi a (1+b) + (1/2)
\log \rho \rceil$$ and $$\eta_{\min}= N_{\min}(a,b,\rho)=\lceil e\pi a
(1-b) + (1/2) \log \rho \rceil.$$

\begin{figure} [ht]
  \centering
  \includegraphics[width=\columnwidth]{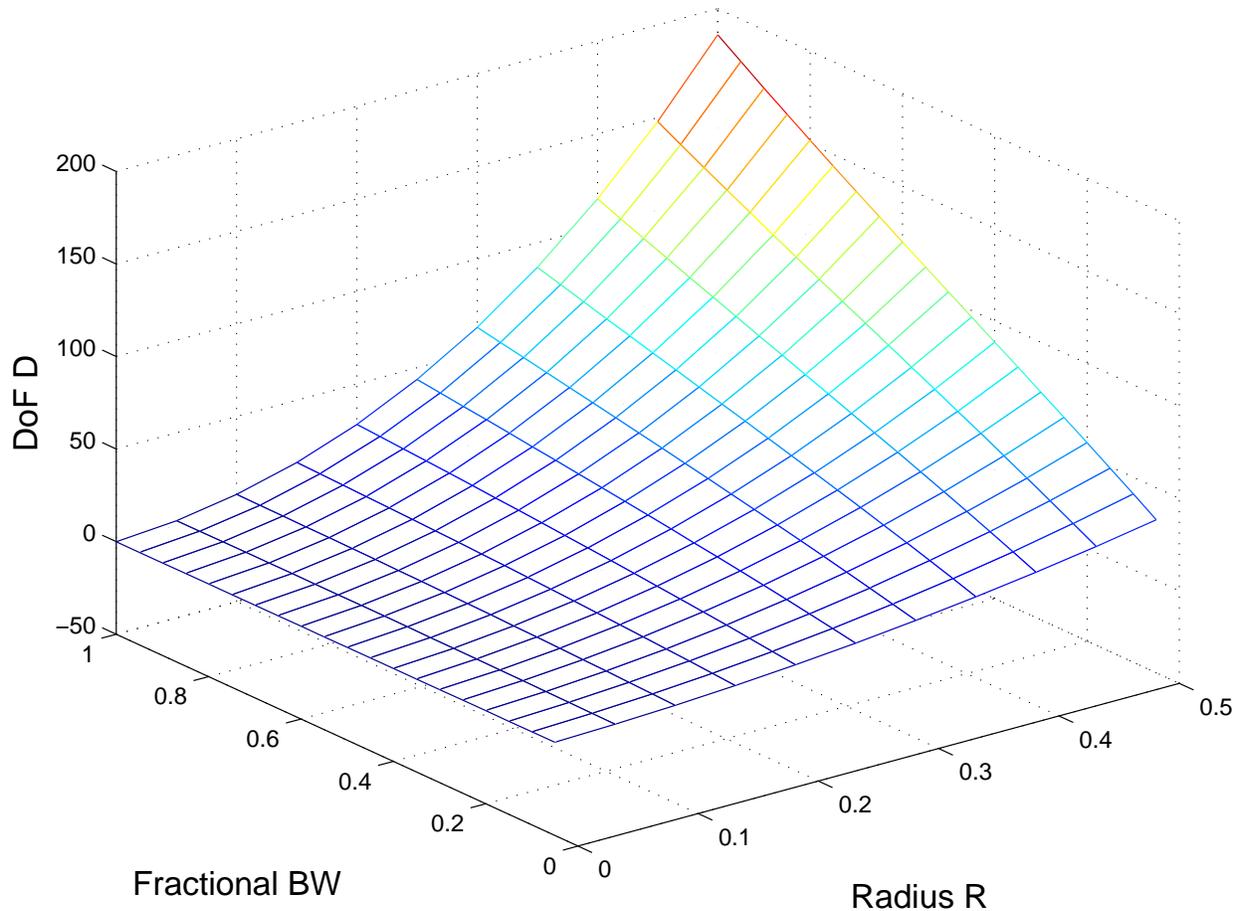}
\caption{ Degrees of Freedom (DoF) $D$ vs. radius $R$ vs. fractional BW at a fixed observation time window ($d=1$) for
$\rho=.5$. Radius, fractional BW and observation time are defined in \eqref{abd}.} \label{STF}
\end{figure}

In Fig. \ref{STF}, degrees of freedom $D$ in \eqref{newD} is plotted
as a function of radius of the spherical region and fractional
bandwidth. It is evident from the figure that for a given
observation time window, there is a sub-quadratic growth in available
degrees of freedom with increasing region size and bandwidth. Note
that we consider a scenario with a small value of $\rho$. Increasing
$\rho$ (which is equivalent of minimizing the affect of noise) we
can achieve higher signal degrees of freedom.

\begin{figure} [ht]
\begin{minipage}[b]{1.0\linewidth}
  \centering
  \centerline{\includegraphics[width=.6\textwidth,height=6.5cm]{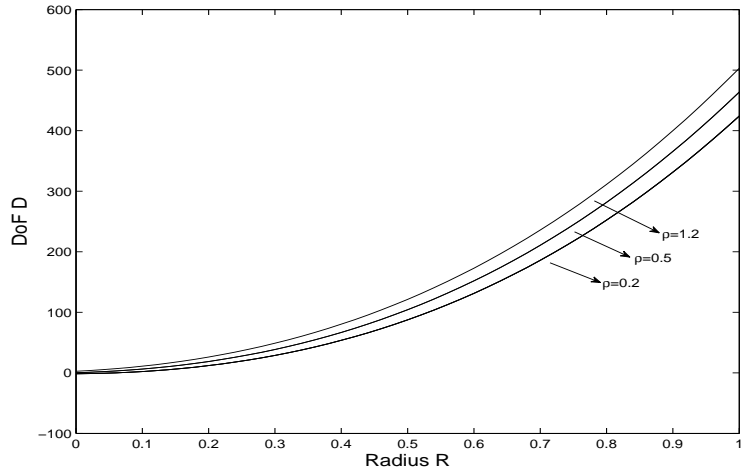}}
%  \vspace{2.0cm}
  \centerline{\small{(a) DoF $D$ vs. radius $R$.} }\medskip
\end{minipage}
\begin{minipage}[b]{1.0\linewidth}
  \centering
  \centerline{\includegraphics[width=.6\textwidth,height=6.5cm]{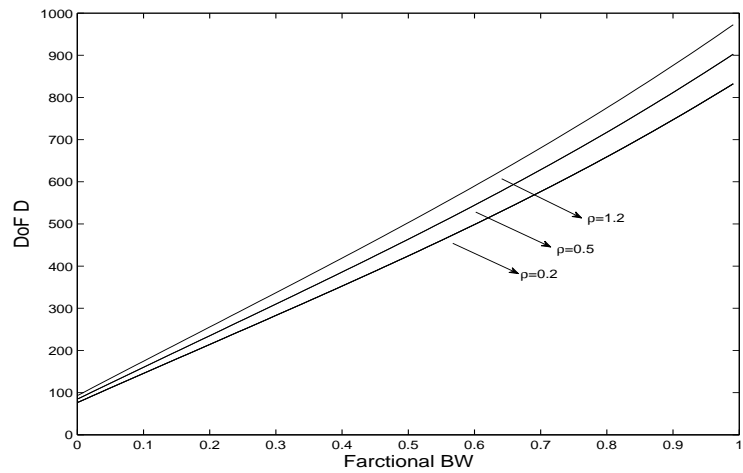}}
%  \vspace{2.0cm}
  \centerline{\small{(b) DoF $D$ vs. fractional BW}}\medskip
\end{minipage}
\medskip \caption{ Degrees of Freedom for different values of $\rho$ at a fixed time window ($d=1$) (a) for a fixed bandwidth ($b=0.5$), (b) for a fixed radius of the region ($a=1$).} \label{rho}
\end{figure}

Next, we portray the affect of SNR on signal degrees of freedom
considering different values of $\rho$. It is evident from Fig.
\ref{rho} that for a given observation time window, increasing the value of
$\rho$, we can obtain a growth in degrees of freedom as a function
of (a) radius of the observable region and (b) bandwidth,
respectively.

\begin{figure} [ht]
\begin{minipage}[b]{1.0\linewidth}
  \centering
  \centerline{\includegraphics[width=.6\textwidth,height=6.5cm]{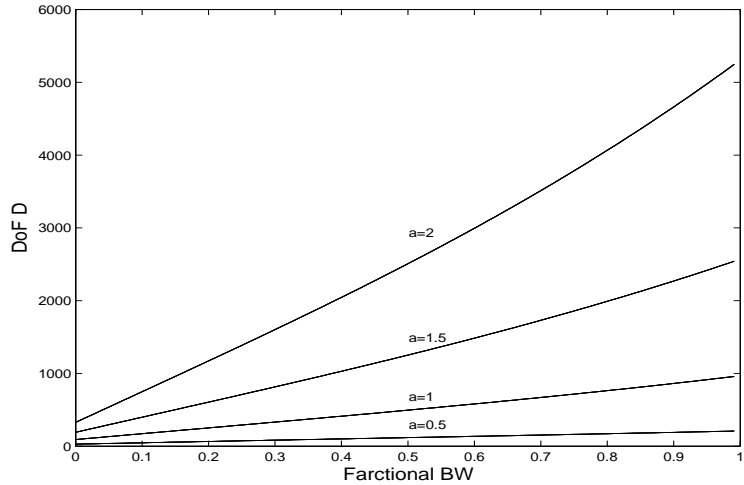}}
%  \vspace{2.0cm}
  \centerline{\small{(a) DoF $D$ vs. fractional BW.} }\medskip
\end{minipage}
\begin{minipage}[b]{1.0\linewidth}
  \centering
  \centerline{\includegraphics[width=.6\textwidth,height=6.5cm]{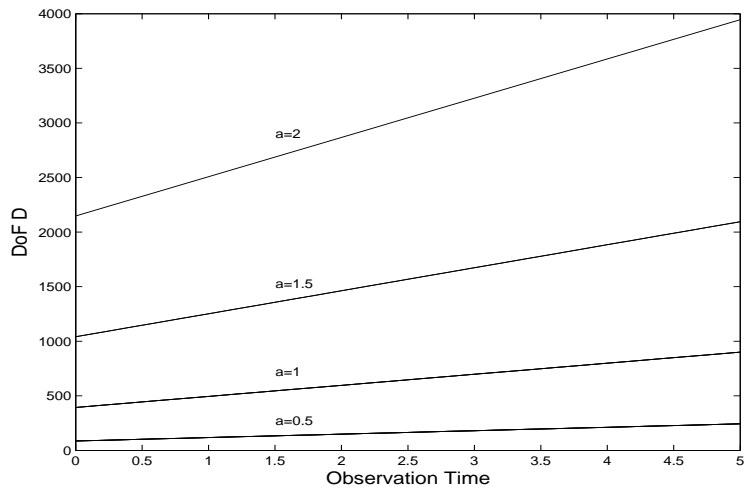}}
%  \vspace{2.0cm}
  \centerline{\small{(b) DoF $D$ vs. observation time}}\medskip
\end{minipage}
\medskip \caption{ Degrees of Freedom  at different radius $R=a\lambda_0$ considering $\rho=1$ (a) for a fixed time
window ($d=1$), (b) for a fixed bandwidth ($b=0.5$).} \label{space}
\end{figure}

We now present the signal degrees of freedom as a function of
bandwidth at a fixed observation time window. The parameter is the radius
of the observable spherical region with $a= \{0.5, 1, 1.5, 2\}$. The
results in Fig. \ref{space} (a) demonstrate that the degrees of
freedom  increases sub-quadratically as the radius of the observable
region increases. \\
On the contrary, it is clear from Fig. \ref{space} (b) that
considering a fixed bandwidth, by increasing the radius of the
observable region, it is possible to obtain a rapid non-linear
growth in the degrees of freedom as a function of observation time.
\\
Note that the two scenarios mentioned above clearly indicate that we
can obtain significantly high signal degrees of freedom only by
increasing the radius of the observable region irrespective of
bandwidth or observation time. This significant growth is also true
if we increase the bandwidth but keep the radius of the region and
observation time constant. \\
From the results, increasing the
frequency or the radius leads to a sub-quadratic growth in the
degrees of freedom. On the other hand, increasing the observation
time window or the SNR does not provide such a significant increase in the
degrees of freedom.

\section{Discussion} \label{insights}

In this section, we elucidate the physical insights of the main
results and attempt to briefly discuss about their applications. We
can make the following comments based on Theorem $4$:
\begin{itemize}
\item From a spatial diversity
perspective, in Shannon's proposed communication model
\cite{shannon49}, wideband signals are encoded in only one spatial
mode or one channel over which information is transmitted. On the
contrary, our proposed model contains spatially diverse wideband
signals that are encoded in a finite number of spatial modes $n$. As a
result, intuitively the available degrees of freedom of any wideband
signal observed over finite spatial and temporal windows should be
Shannon's degrees of freedom result extended form one spatial mode
to $n$ modes. Our derived result clearly comply with this intuition.
The number of modes are the number of independent channels available
to receive information due to the availability of measurements
over a spatial region. This means that each mode or channel has its
own time-frequency degrees of freedom. Spatial diversity, therefore,
provides a number of independent channels over which time-frequency
information can be transmitted.
\item The degrees of freedom result does not agree with the well established result of
evaluating degrees of freedom of spatially diverse wideband signals
as a product of spatial degrees of freedom and time-frequency degrees of freedom
\cite{PoonBrodersenTse,Franceschetti2012}. However, in the
propagation of waves even though space, time and frequency are
separate entities, in spatially diverse wideband transmissions,
space and time as well as space and frequency are strongly coupled,
the results of \cite{PoonBrodersenTse,Franceschetti2012} fail to
show those coupling relationships. On the contrary, our derived
result takes the coupling relations in account.
\item Shannon's work considers broadband transmission over a single channel and shows
that the channel has {\em `time-bandwidth product +1'} degrees of
freedom. On the contrary, in addition to broadband transmission, we
take spatial diversity in account. Therefore, in this work, we
consider broadband transmission over finite $n$ number of channels.
Our analysis indicates that each of these channels has {\em
`effective time-effective bandwidth product +1'} degrees of freedom. This means that
considering spatial diversity, we can capture more information from
broadband transmission. For higher modes, the effective/ usable
bandwidth is less than the measured bandwidth and so not all spatial
modes can covey the same amount of information. Therefore, from Theorem $3$,  for modes $n$ above $N_{\min}$, the {\em
`effective time-effective bandwidth product +1'} decreases as the mode $n$ increases.
\end{itemize}

\subsection{Asymptotic Results}
Let us consider that the threshold is equivalent to the maximum SNR
of the $n^{th}$ mode signal spectrum $\alpha_{nm}(\omega)$, i.e.,
$\gamma={(SNR)}_{\alpha,\max}$. This means that signals below the
frequency $F_n= nc/ (e \pi R)$ are submerged in noise and can not be
detected. As a result, from \eqref{Dspace} we obtain
\begin{IEEEeqnarray}{rCl}\label{asymptotic}
D \leq &&\underbrace{{\left(\lceil e \pi
(F_0+ W) \frac{R}{c}\rceil
+1\right)}^2}_{\text{$D_1$}} +\underbrace{2W\left(T+\frac{2R}{c}\right){\left(\lceil e \pi (F_0- W) \frac{R}{c}\rceil + 1\right)}^2}_{\text{$D_2$}} \nonumber \\ && + \underbrace{W\left(T + \frac{2R}{c}\right)\left[{\left(2e \pi \frac{R}{c}\right)}^2 \left(F_0W - \frac{W^2}{3}\right) + 2e
\pi \frac{R}{c} (2F_0- W)\right]}_{\text{$D_3$}}
\end{IEEEeqnarray}
which yields the following observations:
\begin{itemize}
\item For non-spatially diverse multipath fields $(R=0)$, in \eqref{asymptotic}, $D_1$ reduces to 1,
$D_2$ reduces to $2WT$ and $D_3$ becomes zero. Thus, non-spatially
diverse multipath fields provide $2WT+1$ degrees of freedom.
\item For narrowband wavefields $(W=0)$, in \eqref{asymptotic}, $D_1$ reduces to $( \lceil e \pi F_0 R/c\rceil+1)^2$,
whereas, both $D_2$ and $D_3$ become zero. Hence, there are $(
\lceil e \pi F_0 R/c \rceil+1)^2$ degrees of freedom available in
narrowband wavefields operating at the mid band frequency $F_0$. As
a result, in terms of wavelengths, degrees of freedom available in
narrowband wavefields is $(\lceil e \pi R/\lambda_0 \rceil+1)^2$
where $\lambda_0=c/F_0$ is the wavelength corresponding to the mid band
frequency $F_0$.
\item If any signal observable within a multipath field is representable with only one sample in time domain $(T=0)$, then, by substituting $T=0$ in
\eqref{asymptotic}, we obtain
\begin{IEEEeqnarray}{rCl} \label{T=0}
D&\leq&{\left(\lceil e \pi (F_0+  W)
\frac{R}{c}\rceil +
1\right)}^2 + 4W\frac{R}{c}{\left(\lceil e \pi (F_0- W) \frac{R}{c}\rceil + 1\right)}^2 \nonumber \\ &&
+2W \frac{R}{c} \left[{\left(2e \pi
\frac{R}{c}\right)}^2  \left(F_0W -
 \frac{W^2}{3}\right) + 2e
\pi \frac{R}{c} (2F_0-  W)\right].
\end{IEEEeqnarray}
This equation indicates that even when there is only one
sample available in time domain, spatial diversity influences the
amount of information that can be captured within the observable
region.
\end{itemize}
The derived result \eqref{asymptotic} represents the degrees of
freedom of signals observable within a broadband multipath field
over finite spatial and time windows, assuming the signals are
submerged in noise for frequencies less than the critical frequency
$F_n= nc/ (e \pi R)$ and are not detectable. This result is
consistent with Shannon's time-frequency degrees of freedom result
\cite{shannon49} when we take sample at a single spatial point
$(R=0)$. Further, \eqref{asymptotic} is consistent with the degrees
of result derived in \cite[eqn. 44]{rodParastooTDjones} at
wavelength $\lambda_0$ when we consider narrowband frequency
transmissions $(W=0)$.

\subsection{Applications}
The degrees of freedom result obtained in this paper can be used to provide insights and bounds in the following areas:
\begin{itemize}
\item In the context of spatial broadband communications such as
wireless MIMO communications, this work addresses the fundamental
question of how the spatial degrees of freedom is interrelated to
the time-frequency degrees of freedom. The result provides insights
into gains or losses of degrees of freedom in space and time-frequency analysis.
\item For broadband beamforming the degrees of freedom characterises
the maximum resolution that can be obtained \cite{KrolikSwingler}. The greater the degrees
of freedom the higher the resolution can be obtained. This has particular importance in this area since we have more variables to
work with to perform broadband beamforming. The performance of beamforming in
wireless networks improves with the available degrees of freedom and has been shown in \cite{KeyiChampagne}. In next generation of
wireless communications capable of transmitting high quality video
and audio, array gain is obtained by using broadband beamforming which
exploits the spatial degrees of freedom and the effective bandwidth
of each of these spatial degrees. Our work shows that as more spatial
degrees are exploited for beamforming, for a receive antenna occupying
a limited spatial region, the effective bandwidth of the higher spatial
degrees or modes $n$ are less than the bandwidth and decreases with the mode index $n$.
\item For broadband reception of signals by antennas placed within a given
spatial region, initially linear growth in the degrees of freedom is
obtained with increasing number of antennas. Once the number of antennas
is greater than $(N_{\min}+1)^2$, the increase in degrees of freedom reduces with
each antenna added until number of antennas is equivalent to $(N_{\max}+1)^2$.
After that no gain can be obtained. This is because the wavefield constraint
results in correlations between channels when the number of antennas becomes too large.
\item Interference alignment is a promising new area introduced in the last two decades.
This seeks to solve the spectrum scarcity in wireless communications by utilizing
the available degrees of freedom in space, time and frequency \cite{GomadamCadambeJafar,HuaGengGouJafar12}.
However, in MIMO wireless networks, interference alignment uses the parallel
channels in space offered by spatial degrees of freedom for alignment.
Our derived results can be used to develop an interference alignment scheme
for MIMO wireless networks.  Our results show that optimal signal alignment
needs to consider that not all spatial channels are equal and can place
interference on the spatial channels that have the lower time-frequency
degrees of freedom.  Hence, the interference channels should be placed in
the spatial channels corresponding to the higher spatial modes. Further,
the degrees of freedom analysis of this work provides the maximum
degrees of freedom that can be utilized in these broadband communications with interference.
\item Recently, distributed MIMO communications have seen an increase
in importance due to the popularity of sensor and ad-hoc networks.
Distributed MIMO includes all muti-user communication configurations
where the communications input and outputs are distributed over different users.
Works of \cite{AyferOlivierTse07,FranceschettiMiglioreMinero09,AyferOlivierTse12}
have studied the spatial degrees of freedom for these considering different
channel conditions and showed performance gains. These works looked at only
narrowband transmissions, however, practical wireless transmissions are performed
over a bandwidth. Considering this, our work shows the maximum degrees of freedom
available over space, time and frequency for users in a limited spherical region
cooperating to receive broadband information. Also, we show how the time-frequency
degrees of freedom is distributed over the spatial modes.

\end{itemize}

\section{Conclusion} \label{conclusion}

This paper provides an upper bound to the degrees of freedom of any
signal observed within a band limited multipath wireless field over
finite spatial and temporal windows. This upper bound is obtained characterizing
the multipath field as solution to Helmholtz wave equation
encoded in a finite number of spatial modes. The analysis of the work shows that the effective observation time is
independent of spatial modes and related to the spatial dimension of
the observable region. Further, for broadband transmissions, at each spatial mode,
the observable signals are band limited within an effective frequency
bandwidth and depending on the mode, the effective bandwidth varies from the given frequency bandwidth.
These findings imply that when both
spatial diversity and broadband transmissions are taken in account,
space and time as well as space and frequency cannot be decoupled. \\
The degrees of freedom result derived in this work
takes the coupling relations into account and portrays the interrelation between
spatial degrees of freedom and time-frequency degrees of freedom. From a spatial diversity
perspective, Shannon's proposed communication model considers
wideband signal encoded in only one spatial
mode or one channel over which information is transmitted,
the available degrees of freedom of spatially diverse wideband
signal encoded in finite number of spatial modes $n$ is
Shannon's degrees of freedom result extended form one spatial mode
to $n$ modes. This means that each mode or channel has its
own time-frequency degrees of freedom. \\
We also show that analogous
to time, space can be treated as an information bearing object,
since degrees of freedom increases sub-quadratically as the size of
the observable spatial region increases irrespective of bandwidth or time window.
Further, the derived result portrays how the
degrees of freedom is affected by the acceptable SNR at each spatial
mode.
%Some potential applications of this work includes (\rom{1}) evaluating the optimum number of
%antennas required to maximize performance for users in a limited spherical region
%cooperating to receive broadband information, (\rom{2}) characterizing the
%degrees of freedom of broadband MIMO networks and distributed MIMO networks, (\rom{3}) specifying an efficient
%interference alignment scheme for MIMO networks and (\rom{3}) determining broadband beamforming techniques for next generation
%wireless communication to provide high quality video and audio. A possible extension of
%this paper would be to evaluate the maximum transmission rate users can obtain in cooperative networks.

\appendices
\section{Proof of Theorem $1$} \label{A1}
\begin{IEEEproof} Let $\psi_{nm}(r, t)$ be the inverse Fourier transform of $\Psi_{nm}(r,
\omega)$, then, the inverse Fourier transform of \eqref{signal} is
\begin{IEEEeqnarray}{rCl}\label{OT}
\psi_{nm}(r,t)= {a}_{nm}(t)\ast P_{n}(\frac{tc}{r})
\end{IEEEeqnarray}
where the time domain coefficients ${a}_{nm}(t)$ are the inverse
Fourier transform of $\alpha_{nm}(\omega)$ and the Legendre
polynomials $P_{n}(tc/r)$ are the inverse Fourier transform of
$j_n(\omega r/c)$.

In \eqref{OT}, we can consider $a_{mn}(t)$ as the $n^{th}$ mode time domain signal. Therefore, the $n^{th}$ mode space-time signal $\psi_{mn}(r,t)$ is a convolution between the $n^{th}$ mode time domain signal $a_{mn}(t)$ with the Legendre polynomial $P_n(tc/r)$. The convolution with the Legendre polynomial $P_n(tc/r)$ represents the wavefield constraint and information content in the $n^{th}$ mode space-time signal $\psi_{mn}(r,t)$ is contained in $a_{mn}(t)$. Further, from the definition in \cite[p. 23]{coltonkress98}, Legendre polynomials $P_{n}(tc/r)$ are defined only for $-r/c\leq t \leq r/c$. This characteristic of Legendre polynomials is also evident from Fig. \ref{LP}.
\begin{figure}[htb]
\centering
\includegraphics[width=\columnwidth]{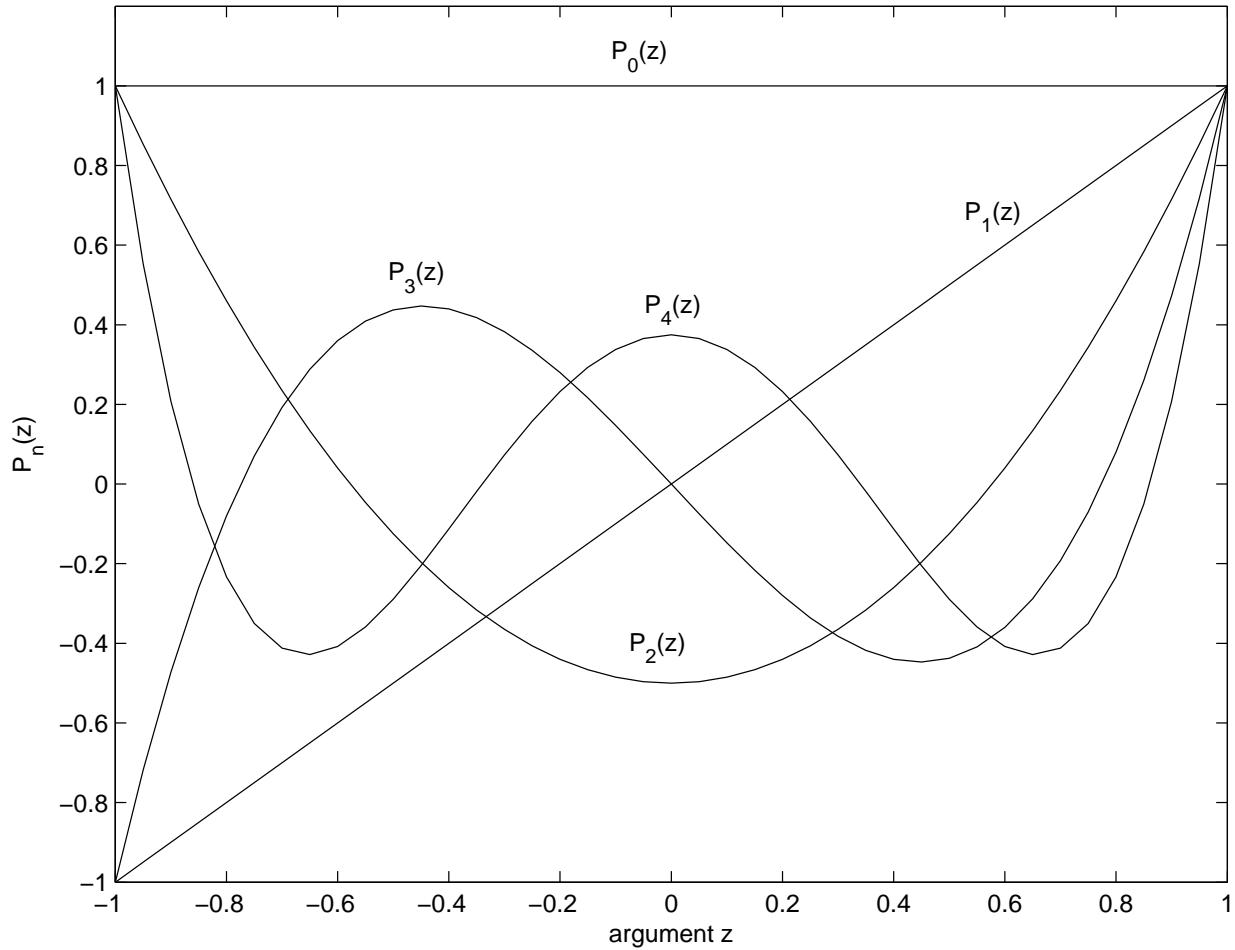}
\caption{Legendre Polynomials $P_n(z)$ for $n= 0, 1, 2, 3, 4$.}
\label{LP}
\end{figure}\\
We observe the $n^{th}$ mode space-time signal $\psi_{nm}(r, t)$ over a time window $[0, T]$ within a spatial window $0 \leq r \leq R$. Therefore, for the given wavefield constraint, at any particular radius $r$, we can observe information in the the $n^{th}$ mode time domain signal $a_{mn}(t)$ over the time window $[-r/c, T+r/c]$. If we consider that the $n^{th}$ mode space-time signal is observed within a sphere of radius $R$, it is possible to capture information about the $n^{th}$ mode time domain signal $a_{nm}(t)$ over a maximum of $T+2R/c$ time interval. Hence, observing the $n^{th}$ mode space-time signal $\psi_{nm}(r, t)$ within a spherical region of radius $R$ over the time window $[0, T]$ is equivalent to observing the $n^{th}$ mode time domain signal $a_{nm}(t)$ over a maximum time window $[-R/c, T+R/c]$.
\end{IEEEproof}

\section{Proof of Theorem $3$} \label{A2}
\begin{IEEEproof} Observable multipath field on the surface of the sphere (at radius
$R$) is
\begin{IEEEeqnarray}{rCl}
\Psi(R, \hat{\textbf{\emph{x}}}, \omega)=
\sum_{n=0}^\infty \sum_{m=-n}^{n} \varPsi_{nm}(R,\omega)
Y_{nm}(\hat{\textbf{\emph{x}}})
\end{IEEEeqnarray}
Applying Parserval's theorem with respect to the spherical harmonics\footnote{Since the spherical harmonics are independent
of each other, we can encode each mode with independent signal
spectrums to achieve the maximum degrees of freedom observed in the
region.}, the average power of the observable multipath field from all propagation directions is
a sum over the average power in the different modes, such that
\begin{IEEEeqnarray}{rCl}
\int_{\mathbb{S}^2} E\{{|\Psi(R, \hat{\textbf{\emph{x}}},
\omega)|}^2\} d\hat{\textbf{\emph{x}}}= \sum_{n=0}^\infty
\sum_{m=-n}^{n} E\{{|\varPsi_{nm}(R,\omega)|}^2\}.
\end{IEEEeqnarray}
Since the noise is independent
of the signal, using \eqref{coefficient} we obtain
\begin{IEEEeqnarray}{rCl} \label{parserval}
E\{{|\varPsi_{nm}(R,\omega)|}^2\}= E\{{|\alpha_{nm}(\omega)|}^2\}
{|j_n\left(\frac{\omega}{c}R\right)|}^2 + E\{{|\nu_{nm}(\omega)|}^2\}.
\end{IEEEeqnarray}
According to Corollary $1$, $E\{{|\nu_{nm}(\omega)|}^2\}= \sigma_0^2$. Therefore,
we can rewrite \eqref{parserval} as
\begin{IEEEeqnarray}{rCl}
E\{{|\varPsi_{nm}(R,\omega)|}^2\}= E\{{|\alpha_{nm}(\omega)|}^2\}
{|j_n\left(\frac{\omega}{c}R\right)|}^2 + \sigma_0^2.
\end{IEEEeqnarray}
From this, signal to noise ratio (SNR) for the $(m,n)^{th}$ mode at frequency $\omega$ is
\begin{IEEEeqnarray}{rCl}
{(SNR)}_{nm} (\omega) &=&
\frac{E\{{|\alpha_{nm}(\omega)|}^2\}{|j_n\left(\frac{\omega}{c}R\right)|}^2
}{\sigma_0 ^2} \label{snrn} \\
&\leq& {(SNR)}_{\alpha, \max} {|j_n\left(\frac{\omega}{c}R\right)|}^2
\label{snr}
\end{IEEEeqnarray}
where  \eqref{snr} follows from \eqref{emax} and \eqref{k}. \\
Next, based on the properties of the Bessel functions \cite[p.
362]{AbramowitzStegun} and the spherical Bessel functions \cite[p.
437]{AbramowitzStegun} and following a few intermediate steps, we can
derive the following bound for the spherical Bessel functions for large $n$
\begin{IEEEeqnarray}{rCl} \label{j}
|j_n(z)| \leq \frac{\sqrt\pi}{2} {\left(\frac{z}{2}\right)}^n \frac{1}{ \Gamma
(n+ 3/2)}, ~~\mbox{$z\geq 0$}
\end{IEEEeqnarray}
where $\Gamma(\cdot)$ is the Gamma function. Therefore,
\begin{IEEEeqnarray}{rCl} \label{snr3}
{(SNR)}_{nm}(\omega)\negthickspace < \negthickspace {(SNR)}_{\alpha, \max} \frac{e}{2 {(2n+1)}^2}\negthickspace {\left(\frac{e
\omega R/c}{2n+1}\right)}^{2n}.
\end{IEEEeqnarray}
This result is obtained using the Stirling lower bound for the Gamma
functions
\begin{IEEEeqnarray}{rCl}
\Gamma(n+3/2) > \sqrt{2\pi e} {\left(\frac{n+1/2}{e}\right)}^{n+1}.
\end{IEEEeqnarray}
Now, using the exponential inequality, $(1+ x/n)^n \leq e^x$ for $n
\neq 0$, we rewrite \eqref{snr3} as
\begin{IEEEeqnarray}{rCl}
{(SNR)}_{nm} (\omega)\leq {(SNR)}_{\alpha,\max} \beta e^{(e \omega R/c-2n)},
\end{IEEEeqnarray}
where $\beta = 2en^2/{(2n+1)}^4$. Since $\beta < 1$, for $n>0$, we have
\begin{IEEEeqnarray}{rCl}
{(SNR)}_{nm} (\omega)< {(SNR)}_{\alpha,\max} e^{(e \omega R/c-2n)}.
\end{IEEEeqnarray}
Note that at the $(m,n)^{th}$ mode, it is not possible to detect signals within the band
of frequencies where the SNR drops below a
certain threshold $\gamma$. Hence, for a frequency to be usable to capture information at the $(m,n)^{th}$ mode,
the SNR must be larger than or equal to the threshold $\gamma$. The
frequency at which the $(SNR)_{n,m}(\omega)$ is at least equal to the threshold $\gamma$ is the critical frequency $F_n$ (where
$\omega_n=2\pi F_n$). Therefore,
\begin{IEEEeqnarray}{rCl}\label{eq:detec}
{(SNR)}_{\alpha,\max} e^{(2\pi e F_n R/c-2n)} = \gamma.
\end{IEEEeqnarray}
This result is easily derived based on the reasoning provided in Section \ref{effectiveBW}. Here, we briefly discuss the reasoning:
as depicted in Fig. \ref{sbessel}, expect for the $0^{th}$ order, spherical Bessel functions show a high pass characteristics. As a result,
for frequencies less then a critical frequency $F_n$, the magnitude of the $n^{th}$ $(n>0)$ order spherical Bessel function
is negligible. Therefore, at each spatial mode $n>0$, for frequencies less than the critical frequency $F_n$, it is not possible to maintain the SNR at least
equal to the threshold $\gamma$. \\
Now, making $F_n$ the subject of the formula in \eqref{eq:detec} yields
\eqref{cf}. This means that for spatial modes
$n>0$, signals below frequency $F_n$ \eqref{cf} are not detectable
since \eqref{eq:detec} will not be satisfied. Observe that for any
particular mode $n(>0)$, if $F_n>F_0-W$, the effective bandwidth of
that mode is $F_0+W-F_n$. In addition, if $F_n>F_0+W$, the effective
bandwidth of this mode and modes above this is zero. It should also
be noted that for a fixed value of radius, $j_0(z)$ is active within
the frequency range $[0, \infty)$ as depicted in Fig \ref{sbessel},
hence, effective bandwidth of the $0^{th}$ mode is $2W$. These
arguments are written mathematically as \eqref{BW}.
\end{IEEEproof}

\section{Proof of the orthogonality of
the functions $\phi_{\ell}(t)$} \label{A3}
\begin{IEEEproof} Since
\begin{IEEEeqnarray}{rCl}
\phi_{\ell}(t) =e^{j2\pi
W_{0n}(t-\frac{\ell}{W_n})} \frac{\sin\pi
W_n(t-\frac{\ell}{W_n})}{\pi
W_n(t-\frac{\ell}{W_n})},
\end{IEEEeqnarray}

\begin{IEEEeqnarray}{rCl} \label{timebasis}
\int_{-\infty}^{\infty} \phi_\ell(t) \phi^{\ast}_{\ell'}(t) dt &=& \negthickspace \int_{-\infty}^{\infty} \negthickspace e^{j2\pi
W_{0n}(t-\frac{\ell}{W_n})} \frac{\sin\pi
W_n(t-\frac{\ell}{W_n})}{\pi
W_n(t-\frac{\ell}{W_n})} \nonumber  e^{-j2\pi
W_{0n}(t-\frac{{\ell}^\prime}{W_n})} \frac{\sin\pi
W_n(t-\frac{\ell^\prime}{W_n})}{\pi
W_n(t-\frac{\ell^\prime}{W_n})} dt \nonumber \\
&=& \kappa \int_{-\infty}^{\infty} \frac{\sin\pi
W_n(t-\frac{\ell}{W_n})}{\pi
W_n(t-\frac{\ell}{W_n})}  \frac{\sin\pi
W_n(t-\frac{\ell^\prime}{W_n})}{\pi
W_n(t-\frac{\ell^\prime}{W_n})} dt \nonumber \\
&=& \left\{ \begin{array}{rl}
   0 &\mbox{$\ell\neq \ell^\prime$} \\
   \frac{1}{W_n} &\mbox{$\ell=\ell^\prime$}
       \end{array} \right.
\end{IEEEeqnarray}
where $\kappa= e^{-j2\pi (\ell-\ell^\prime)
W_{0n}/W_n}$ and for $\ell=\ell^\prime$, $\kappa=1$. Here, \eqref{timebasis} is derived using the fact that \cite[eqn. 11]{shannon49}
\begin{IEEEeqnarray}{rCl}
 &\int_{-\infty}^{\infty}& \frac{\sin\pi
W_n(t-\frac{\ell}{W_n})}{\pi
W_n(t-\frac{\ell}{W_n})}  \frac{\sin\pi
W_n(t-\frac{\ell^\prime}{W_n})}{\pi
W_n(t-\frac{\ell^\prime}{W_n})} dt \nonumber \\
&=& \left\{ \begin{array}{rl}
   0 &\mbox{$\ell\neq \ell^\prime$} \\
   \frac{1}{W_n} &\mbox{$\ell=\ell^\prime.$}
       \end{array} \right.
\end{IEEEeqnarray}
Thus, the functions $\phi_{\ell}(t)$ are orthogonal over time.
\end{IEEEproof}

\section{Proof of Theorem $4$} \label{A4}
\begin{IEEEproof}
We start from \eqref{eq:D} and by expanding the sum, we obtain
\begin{IEEEeqnarray}{rCl} \label{sum}
D= \sum_{n=0}^{N_{\max}}(2n+1) + T_{eff} \sum_{n=0}^{N_{max}}(2n+1)
W_n.
\end{IEEEeqnarray}
Substituting \eqref{TS} and \eqref{BW} in \eqref{sum} yields
\begin{IEEEeqnarray}{rCl}\label{eq:D1}
 D &=& \sum_{n=0}^{N_{\max}}(2n+1)+ 2W
 (T+\frac{2R}{c})\sum_{n=0}^{N_{\min}}(2n+1) +
(T+\frac{2R}{c})\sum_{n>N_{\min}}^{N_{\max}}(2n+1)(F_0+W-F_n) \nonumber \\
&=& D_1+ 2W (T+\frac{2R}{c}) D_2+(T+\frac{2R}{c}) D_3
\end{IEEEeqnarray}
%%%%%%
considering,
\begin{IEEEeqnarray}{rCl}\label{d1}
 D_1 = \sum_{n=0}^{N_{\max}}(2n+1),
\end{IEEEeqnarray}
\begin{IEEEeqnarray}{rCl}\label{d2}
 D_2 = \sum_{n=0}^{N_{\min}}(2n+1)
 \end{IEEEeqnarray}
 and
\begin{IEEEeqnarray}{rCl}\label{d3}
 D_3 = \sum_{n>N_{\min}}^{N_{\max}}(2n+1)(F_0+W-F_n).
\end{IEEEeqnarray}
Now, using the sum of the first $p$ odd numbers, $\sum_{p=0}^{P} (2p+1)={(P+1)}^2$, we can rewrite \eqref{d1} and \eqref{d2} as
\begin{IEEEeqnarray}{rCl}\label{d1s}
 D_1 = {(N_{\max}+1)}^2
\end{IEEEeqnarray}
and
\begin{IEEEeqnarray}{rCl}\label{d2s}
 D_2 = {(N_{\min}+1)}^2.
 \end{IEEEeqnarray}
Further, replacing $F_n$ by \eqref{cf} in \eqref{d3} and following a few
intermediate steps, we obtain
\begin{IEEEeqnarray}{rCl}\label{d3s}
D_3 \leq 2W  \Big[2{\left(e\pi \frac{R}{c}\right)}^2
\left(F_0W-\frac{1}{3}W^2\right) \negthickspace +\negthickspace
e\pi\frac{R}{c}\negthickspace \left(2F_0
\negthickspace - \negthickspace W \right)\negthickspace  +
\log\left(\frac{{(SNR)}_{\alpha,\max}}{\gamma}\right)\negthickspace
\left( \negthickspace e\pi W \frac{R}{c}\negthickspace +
\negthickspace 1 \negthickspace \right)\Big].
 \end{IEEEeqnarray}
Finally, substituting \eqref{d1s}, \eqref{d2s} and \eqref{d3s} in \eqref{eq:D1}, we deduce the upper bound on the signal degrees of freedom \eqref{Dspace}.
\end{IEEEproof}

\bibliographystyle{IEEEtran}
%\bibliography{ref}


\end{document}